%% file: template.tex
\documentclass{Interspeech}

\interspeechcameraready

\title{On the Language and Gender Biases in PSTN, VoIP and Neural Audio Codecs}

\author[affiliation={1,2}]{Kemal}{Altwlkany}
\author[affiliation={2}]{Amar}{Kuric}
\author[affiliation={3}]{Emanuel}{Lacic}

\affiliation{Faculty of Science}{University of Sarajevo}{Bosnia and Herzegovina}
\affiliation{Infobip}{Sarajevo}{Bosnia and Herzegovina}
\affiliation{Infobip}{Zagreb}{Croatia}
\email{\{kemal.altwlkany, amar.kuric, emanuel.lacic\}@infobip.com}


\usepackage{comment}

\begin{document}

\maketitle

\begin{abstract}
    


    
In recent years, there has been a growing focus on fairness and inclusivity within speech technology, particularly in areas such as automatic speech recognition and speech sentiment analysis. When audio is transcoded prior to processing, as is the case in streaming or real-time applications, any inherent bias in the coding mechanism may result in disparities. This not only affects user experience but can also have broader societal implications by perpetuating stereotypes and exclusion. Thus, it is important that audio coding mechanisms are unbiased. In this work, we contribute towards the scarce research with respect to language and gender biases of audio codecs. By analyzing the speech quality of over 2 million multilingual audio files after transcoding through a representative subset of codecs (PSTN, VoIP and neural), our results indicate that PSTN codecs are strongly biased in terms of gender and that neural codecs introduce language biases.

\end{abstract}

\input{text/introduction}

\input{text/method}

\input{text/results}

\input{text/discussion}
\input{text/conclusion}

\bibliographystyle{IEEEtran}
\bibliography{mybib}

\end{document}

%% file: text/introduction.tex
\section{Introduction}

Audio coding has been present for more than 60 years and recent studies suggest that audio and video streaming account for the majority of Internet traffic (i.e., 82\%) \cite{gersho2002advances,defossezhigh}. The terms \textit{audio coding}, \textit{audio codec} and \textit{audio compression} are nowadays used interchangeably since all terms share the same main premise: reducing or compressing an input audio signal while maintaining an acceptable level of fidelity and quality of the original signal.

Traditional audio codecs such as G.711 \cite{g711} and G.729 \cite{g729} were originally designed for use in telephony - today still referred to as the public switched telephone network (PSTN). The audio codecs here operate deterministically, typically taking a small chunk of the input audio sampled at a predefined rate (e.g., \SI{10}{\milli\second} of audio sampled at \SI{8}{\kilo\hertz}) and compressing it into a smaller number of bits. These bits are streamed over the network and expanded at the receiving end to restore the original signal. More recently, the popularization of the Internet has seen voice over IP (VoIP) telephony technologies become a crucial part in human communication. This resulted in the development of audio codecs based on linear predictive coding (LPC) and modified discrete cosine transform (MDCT), such as SILK, EVS \cite{dietz2015overview} and most notably Opus \cite{valin2012definition}, one of the codecs used by millions of daily users in applications such as Microsoft Teams, Google Meet, YouTube and others \cite{zeghidour2021soundstream}.
Recent advances in deep learning have seen the development of neural-based audio codecs, such as EnCodec \cite{defossezhigh}, SoundStream \cite{zeghidour2021soundstream} and Descript \cite{kumar2024high}. These models are trained to encode music and speech at varying bitrates with streaming capabilities.

\subsection{Related work}

Most work on fairness, inclusivity and biases of speech technologies has revolved around automatic speech recognition. This is not surprising given the high demand and applicability of these technologies \cite{tatman2017gender,tatman17_interspeech,koenecke2020racial,mengesha2021don,liu2022towards,dheram22_interspeech}. These studies mostly analyze whether automatic speech recognition (ASR) systems are fair regardless of gender, age and ethnicity/race. The concept of \textit{language bias} is not applicable to ASR systems as the performance of an ASR system on a specific language is a target metric and is improved by training the ASR system on data in that given language. Several studies analyzed the gender disparity between emotion recognition from speech, suggesting favorable accuracy for male samples compared to female samples \cite{gorrostieta19_interspeech,luitel2024investigating}. The authors of \cite{holub2023human} also show that disparities in speech quality transmission based on gender exist.

In \cite{lai24_interspeech}, the authors consider the effect of voice quality (i.e. timbre component of speech) on ASR accuracy, primarily analyzing the effects of \textit{creaky} voice on ASR performance. Furthermore, \cite{xu24j_interspeech} analyzes the disparity between acoustic measures extracted from high-quality recordings of voice and audio transcoded via VoIP, specifically using Opus \cite{valin2012definition}. 
Most relevant to our work is the study of Muller et al. \cite{muller24c_interspeech} in which the authors evaluate the quality of various audio codecs and provide a partial per-gender assessment of the data. However, a limitation there is that it was conducted in French only and did not include any traditional PSTN codecs.

To the best of our knowledge, our paper is the first study regarding a per-language analysis of audio codecs, especially one which encompasses PSTN, VoIP and neural codecs.

\subsection{Motivation}


We argue that even though there is research on the fairness and inclusivity of ASR systems, its importance regarding audio codecs should not be disregarded. These codecs are often the first step before audio is processed by speech recognition and analysis models. Popular speech-to-text service providers such as Amazon\footnote{https://aws.amazon.com/transcribe/}, Google\footnote{https://cloud.google.com/speech-to-text}, Microsoft\footnote{https://azure.microsoft.com/en-us/products/ai-services/ai-speech}, and OpenAI\footnote{https://platform.openai.com/docs/guides/realtime} among others offer these services over the Internet, thereby transcoding audio during the process. ASR is often directly incorporated into video conferencing technologies, which as we mentioned relies on VoIP codecs.

Voice providers such as Infobip\footnote{https://www.infobip.com/docs/api/channels/voice}, Telnyx\footnote{https://telnyx.com/products/voice-api} and Twilio\footnote{https://www.twilio.com/docs/voice} provide speech-to-text on live VoIP calls and often neglected PSTN calls. As such, an objective of this paper is to highlight the need to include traditional PSTN audio codecs in the evaluation of audio codecs, given their strong presence in telephony and speech technology. To support this claim, we share data provided to us by Infobip in Table \ref{tab:pstn-codec-count}. The data was collected during the second half of 2024 from 41 data centers located on all continents which process worldwide PSTN traffic. The large number of calls created only through Infobip's platform indicates that traditional PSTN codecs are still in use and we decided to include them in our study. 
Although this does not include PSTN traffic from other vendors and telecommunication platforms, nor PSTN traffic that was not transcoded by Infobip, these statistics still provide a powerful and insightful representation of the current state of the telecommunications industry.
As such, this paper explores the research question whether PSTN, VoIP, or neural audio codecs introduce language or gender bias, and if so, to what extent.


%% file: text/method.tex
\section{Method}

When investigating the language and gender biases of audio codecs, there are several criteria that should be addressed when selecting the data and audio codecs on which to conduct such an analysis.

\subsection{Data}


First, the previously-mentioned neural codecs are trained on some of the most popular and diverse speech datasets available.
Both EnCodec \cite{defossezhigh} and Descript \cite{kumar2024high} were trained on the Common Voice dataset \cite{ardila2020common} and clean speech segments from DNS Challenge 4 \cite{dubey2024icassp} (additionally DAPS \cite{mysore2014can} and VCTK \cite{kumar2024high} for Descript). These datasets are not suitable for our analysis as they would yield an unfair comparison against other codecs and could hide the actual biases of these models.
The second issue is that we need data which is diverse in terms of speaker language and gender, which by itself is already limitedly available.
Finally, the third concern that needs to be addressed is regarding the usage of audio recordings that were already compressed using a lossy compression format (e.g., mp3). Although this would not pose an issue for some speech technology applications, here we are directly comparing the capability of codecs to compress audio while maintaining high fidelity of the original signal and remaining unbiased. As to the best of our knowledge, we are not aware of any reports about the impact of already compressed audio on the performance of speech technology in general, especially audio codecs, in this paper we focus our investigation on audio data available in lossless format (e.g. wave or flac). While performing an analysis whether lossy compressed audio formats influence the results of speech technology would be beneficial, it exceeds the scope of this paper and we suggest considering it for a future study.

Given the above-mentioned constraints, we opt for VoxForge \cite{VoxForge} to perform the language analysis. VoxForge is a highly multilingual dataset, containing recordings in $17$ different languages. In total, we have around $170$ thousand audio recordings, totaling $3,656$ hours of speech data. The median and mean duration of the audio recordings are $5.5$ and $5.86$ seconds, respectively. 
Since VoxForge does not provide gender or age-related metadata, we perform the gender and age analysis using DARPA TIMIT \cite{garofolo1993darpa}. This dataset provides recordings only in English, but includes $630$ different speakers from $8$ major dialect regions of the United States. The dataset includes metadata such as speaker gender, age, race, dialect region and education level.

\subsection{Audio coding}

In this work, we transcode audio from the previously mentioned multilingual dataset of several native speakers through PSTN, VoIP and neural codecs. To assess audio quality, we compare audio before and after transcoding using an objective, intrusive audio quality metric. That is, we use ViSQOL \cite{chinen2020visqol} as a metric to assess perceived audio quality. The pretrained model is open source and publicly available\footnote{https://github.com/google/visqol}. We used ViSQOL in speech mode (\textit{v3.3.3} - latest version available at the time of writing).

\vspace{2mm}
\noindent
\textbf{Traditional audio codecs.}
As reported in Table \ref{tab:pstn-codec-count}, PCMA (G.711 A-law) and PCMU (G.711 $\mu$-law) \cite{g711} account for more than 99\% of PSTN traffic facilitated by Infobip. We thus include both PCMA and PCMU variants of G.711 in our analysis. We also include G.729A \cite {g729} despite its relatively low usage compared to G.711, as it is the most used low-bitrate PSTN audio codec (i.e., \SI{8}{kbps}) and will allow for a more fair comparison to low-bitrate versions of VoIP and neural codecs.

\vspace{2mm}
\noindent
\textbf{VoIP audio codecs.}
For VoIP, we select Opus \cite{kumar2024high}, which supports bitrates between \SI{6}{} and \SI{510}{kbps}. We perform the transcoding at the following bitrates: \SI{6}{}, \SI{8}{}, \SI{12}{} and \SI{24}{kbps}. These bitrates were chosen for two reasons: (1) to have bitrates comparable with those offered by neural codecs and G.729A; and (2) they are recommended as the "sweet spots" for narrowband and wideband speech \cite{rfc6716}.



\vspace{2mm}
\noindent
\textbf{Neural audio codecs.}
We focus on EnCodec \cite{defossezhigh} and Descript \cite{kumar2024high} as representatives of neural codecs. With EnCodec, we perform the transcoding at variable bitrates; \SI{3}{}, \SI{6}{}, \SI{12}{} and \SI{24}{kbps}. With Descript, we use the \SI{24}{\kilo\hertz} model. Although the authors report a bitrate of \SI{8}{kbps}, we were not able to reproduce these results - the obtained bitrates were 8 times greater in some cases. After some investigation, we found that we are not alone in experiencing these misalignments, as other Descript users have reported similar issues on the public official GitHub repository referenced in the original Descript paper\footnote{https://github.com/descriptinc/descript-audio-codec/issues/73}. We reached out to the authors, and while awaiting their reply, opted not to entirely exclude Descript from our experiments. We included the results from the \SI{8}{kbps} configuration, but advise readers to interpret these results cautiously.

\input{tables/pstn_codec_usage}

\input{tables/visqol_languages}


%% file: tables/pstn_codec_usage.tex
\begin{table}[!t]
\caption{Number of PSTN calls created  during the second half of year 2024 through Infobip's platform with PSTN codec count.}
\resizebox{1\linewidth}{!}{%
\begin{tabular}{l|c|c|c}
Codec           & G.711 (PCMA)                       & G.711 (PCMU)                       & G.729A                        \\ \hline
Number of calls & \multicolumn{1}{r|}{2,579,290,920} & \multicolumn{1}{r|}{1,701,341,847} & \multicolumn{1}{r}{2,313,558}
\end{tabular}
}
\label{tab:pstn-codec-count}
\end{table}

%% file: tables/visqol_languages.tex
\begin{table*}[!ht]
\caption{ViSQOL scores grouped per codec and bitrate for each language, computed within a 99\% confidence interval.}
\resizebox{\linewidth}{!}{%
\begin{tabular}{l|r|r|r|rrrr|rrrr|r}
           & \multicolumn{1}{c|}{G.711A} & \multicolumn{1}{c|}{G.711$\mu$} & \multicolumn{1}{c|}{G.729A} & \multicolumn{4}{c|}{Opus}                                                                        & \multicolumn{4}{c|}{EnCodec}                                                                     & \multicolumn{1}{c}{Descript} \\ 
           & \multicolumn{1}{c|}{64}      & \multicolumn{1}{c|}{64}       & \multicolumn{1}{c|}{8}      & \multicolumn{1}{c}{6} & \multicolumn{1}{c}{8} & \multicolumn{1}{c}{12} & \multicolumn{1}{c|}{24} & \multicolumn{1}{c}{3} & \multicolumn{1}{c}{6} & \multicolumn{1}{c}{12} & \multicolumn{1}{c|}{24} & \multicolumn{1}{c}{8}        \\ \hline
Dutch      & $2.92\pm.007$                    & $3.02\pm.009$                    & $2.45\pm.006$                  & $2.50\pm.004$            & $2.62\pm.006$            & $3.34\pm.015$             & $3.68\pm.016$              & $2.61\pm.006$            & $2.86\pm.008$            & $3.14\pm.012$             & $3.32\pm.014$              & $4.55\pm.007$                   \\
English    & $2.99\pm.003$                    & $3.03\pm.003$                    & $2.43\pm.002$                  & $2.47\pm.002$            & $2.59\pm.002$            & $3.32\pm.005$             & $3.68\pm.006$              & $2.61\pm.002$            & $2.92\pm.003$            & $3.28\pm.004$             & $3.53\pm.005$              & $4.52\pm.002$                   \\
German     & $2.83\pm.003$                    & $2.85\pm.004$                    & $2.44\pm.002$                  & $2.47\pm.002$            & $2.55\pm.002$            & $3.28\pm.007$             & $3.62\pm.008$              & $2.63\pm.003$            & $2.94\pm.005$            & $3.30\pm.007$             & $3.52\pm.008$              & $4.56\pm.003$                   \\ \hline
French     & $2.97\pm.004$                   & $3.02\pm.006$                    & $2.44\pm.003$                  & $2.49\pm.003$            & $2.62\pm.004$            & $3.41\pm.009$             & $3.80\pm.010$              & $2.67\pm.003$            & $3.01\pm.005$            & $3.40\pm.006$             & $3.66\pm.007$              & $4.54\pm.004$                   \\
Italian    & $3.02\pm.006$                   & $3.05\pm.008$                    & $2.46\pm.004$                  & $2.50\pm.004$            & $2.62\pm.006$            & $3.37\pm.013$             & $3.70\pm.016$              & $2.68\pm.005$            & $3.04\pm.008$            & $3.42\pm.010$             & $3.67\pm.011$              & $4.51\pm.006$                   \\ 
Portuguese & $3.06\pm.012$                   & $3.10\pm.016$                    & $2.42\pm.010$                  & $2.46\pm.007$            & $2.57\pm.010$            & $3.26\pm.022$             & $3.62\pm.024$              & $2.60\pm.007$            & $2.93\pm.013$            & $3.32\pm.018$             & $3.57\pm.021$              & $4.53\pm.009$                   \\
Spanish    & $2.96\pm.004$                   & $3.00\pm.006$                    & $2.45\pm.003$                  & $2.50\pm.003$            & $2.63\pm.008$            & $3.46\pm.009$             & $3.85\pm.010$              & $2.70\pm.003$            & $3.07\pm.005$            & $3.46\pm.008$             & $3.70\pm.008$              & $4.56\pm.004$                   \\ \hline
Greek      & $3.04\pm.017$                   & $3.08\pm.024$                    & $2.43\pm.016$                  & $2.48\pm.010$            & $2.62\pm.016$            & $3.40\pm.037$             & $3.77\pm.043$              & $2.65\pm.011$            & $3.02\pm.019$            & $3.42\pm.027$             & $3.67\pm.031$              & $4.58\pm.014$                   \\
Russian    & $2.99\pm.008$                   & $3.03\pm.012$                    & $2.44\pm.006$                  & $2.49\pm.006$            & $2.63\pm.008$            & $3.38\pm.017$             & $3.77\pm.018$              & $2.62\pm.005$            & $2.95\pm.010$            & $3.31\pm.014$             & $3.55\pm.016$              & $4.53\pm.008$                   \\ 
Turkish    & $3.13\pm.018$                   & $3.19\pm.027$                    & $2.46\pm.019$                  & $2.52\pm.016$            & $2.70\pm.023$            & $3.47\pm.039$             & $3.89\pm.042$              & $2.68\pm.012$            & $3.02\pm.021$            & $3.39\pm.027$             & $3.62\pm.030$              & $4.58\pm.013$                   \\
Other      & $2.96\pm.014$                   & $3.02\pm.019$                    & $2.48\pm.008$                  & $2.52\pm.009$            & $2.66\pm.014$            & $3.52\pm.029$             & $3.91\pm.032$              & $2.68\pm.010$            & $3.03\pm.017$            & $3.41\pm.023$             & $3.64\pm.025$              & $4.60\pm.011$                  
\end{tabular}
}
\label{tab::visqol-languages}
\end{table*}

%% file: text/results.tex
\subsection{Experimental setup}
After transcoding each uncompressed wave file from VoxForge into our set of desired codecs and bitrates, we obtained $2,061,600$ audio files which we use for language analysis. Similarly, to analyze gender bias, we transcode each uncompressed wave file from the DARPA-TIMIT corpus \cite{garofolo1993darpa} with respect to the previously defined set of codecs and bitrates with which we obtain a total of $75,600$ audio files. We evaluated the audio quality of all files using ViSQOL.








\input{tables/anova_t_test_language_group}

\section{Results}
In Table \ref{tab::visqol-languages} we report the mean ViSQOL score computed at a 99\% confidence interval using the bootstrap approach \cite{tibshirani1993introduction} aggregated by language, codec and bitrate. We explicitly report only those languages that contain at least $10,000$ audio files, while we aggregate all other languages into a group called \textit{Other}. Our experiments verified well-known and expected results: in terms of ViSQOL, VoIP codecs at lower bitrates highly outperform traditional PSTN codecs which operate at higher bitrates. The same applies to neural codecs. An increase in the bitrate results in a better ViSQOL score for a given codec.

\subsection{Language bias}
With respect to the results reported in Table \ref{tab::visqol-languages} , we performed an N-way analysis of variance (ANOVA) with the independent variables being codec, bitrate and language and report a statistically significant result in terms of ViSQOL ($p < 0.0001$). Additionally, we performed a post hoc one-way ANOVA considering only language as an independent variable; we analyze the data per each codec and bitrate separately to determine whether language alone has an effect on audio quality. Since this required multiple comparisons, we applied the Benjamini-Hochberg (BH) procedure \cite{benjamini1995controlling} to adjust the $p$-value. After correction, the $p$-value for each codec at each bitrate is $p < 0.0001$, indicating that the transcoded audio quality is dependent on the respective language.

\begin{figure}[!t]
  \centering
  \includegraphics[width=\linewidth]{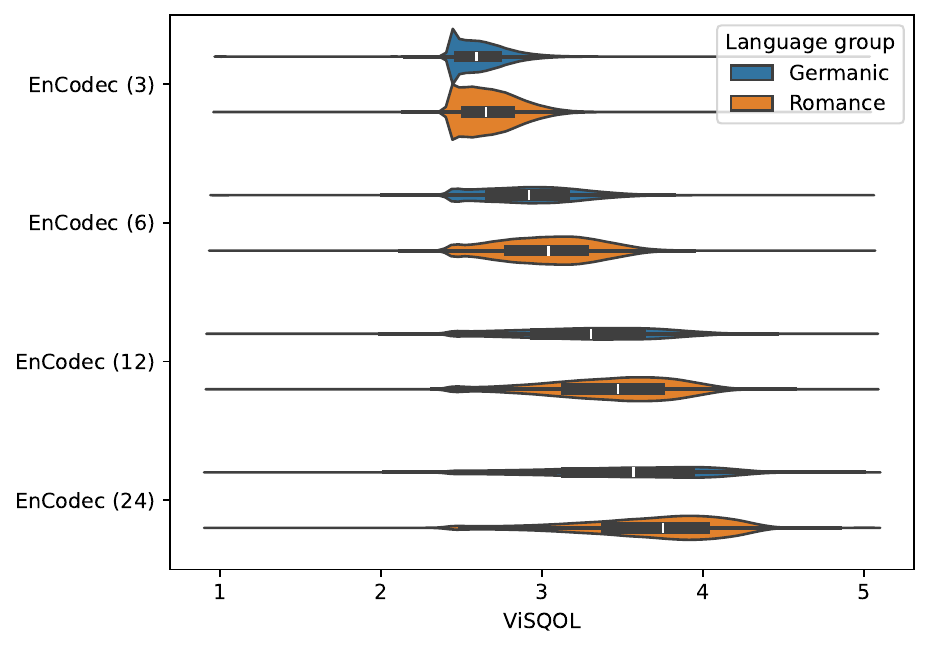}
  \caption{Grouped Germanic and Romance languages that show the distribution of ViSQOL scores (x-axis) for EnCodec at 3, 6, 12 and 24 kbps. The distribution is more spread at higher bitrates and is more spread for Germanic languages. Regardless of bitrate, the mean for Romance languages is higher than that for Germanic languages.}
  \label{fig:boxplot-language-groups}
\end{figure}

\vspace{2mm}
\noindent
\textbf{Germanic and Romance language groups.}
We further analyzed whether there is a difference between Germanic languages (Dutch, English and German) and Romance languages (French, Italian, Portuguese and Spanish). Here we performed a one-way ANOVA and t-test for each codec and compare the two language groups. The effect sizes partial eta squared ($\eta^{2}$) and Cohen's $d$ are reported in Table \ref{tab::anova-t-test-language-groups} and are interpreted as in \cite{ellis2010essential}. Statistically significant results were obtained for each codec and bitrate (after adjustment using BH). The results of ANOVA and the t-test are aligned; only for EnCodec we observe a small to medium effect size both in terms of ANOVA ($\eta^{2} > 0.01$) and the t-test ($d > 0.3$). Thus, we visualize the distribution of ViSQOL scores per language group for EnCodec in Figure \ref{fig:boxplot-language-groups}.

\begin{figure}[!h]
  \centering
  \includegraphics[width=\linewidth]{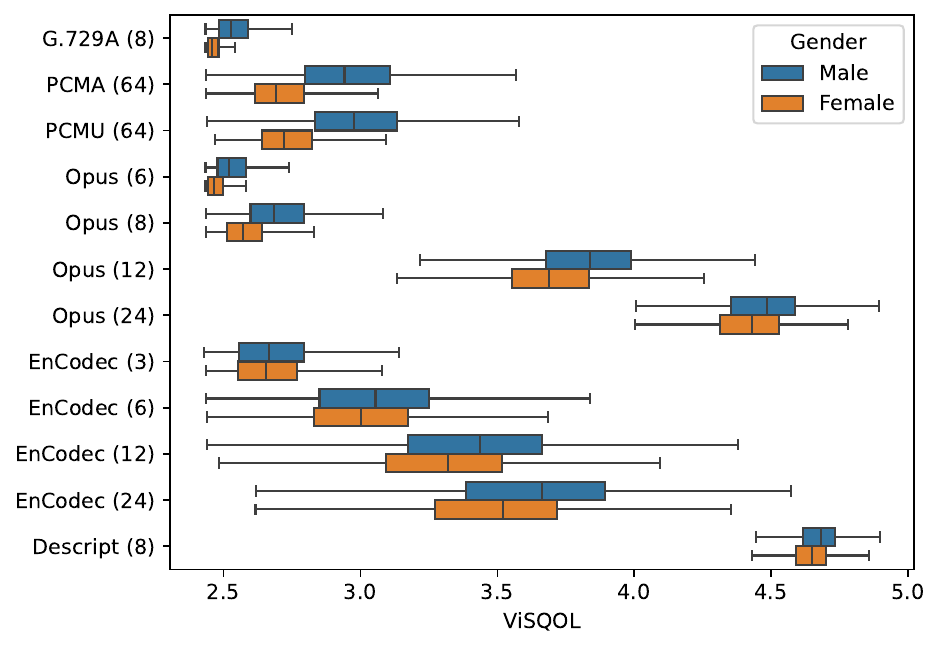}
  \caption{ViSQOL scores (x-axis) grouped by gender for each analyzed codec and bitrate (labels on y-axis). From the plot it is clear that a disparity between ViSQOL scores for male and female speech exists; for PSTN codecs there is no overlap between the 3rd quartile of female speech and 1st quartile of male speech.}
  \label{fig:boxplot-gender}
  \vspace{-4mm}
\end{figure}

\subsection{Gender bias}

In order to assess the bias towards gender, in Table \ref{tab::anova-t-test-gender} we report the experiments using one-way ANOVA and a t-test for each codec and bitrate separately. The $p$-values are again corrected using the Benjamini-Hochberg procedure to adjust for multiple comparisons. We observe statistically significant results for each combination of codec and bitrate, with the highest values of Cohen's $d$ and $\eta^{2}$ obtained for PSTN codecs. The results indicate that audio codecs are biased towards male voices compared to female ones. For VoIP codecs, a medium to strong effect size is observable (e.g. Opus for \SI{6}{kbps}: $\eta^{2} = 0.121$, $d = 0.804$), while neural codecs exhibit an effect size ranging from almost neglible to small (e.g. EnCodec for \SI{6}{kbps}: $\eta^{2} = 0.007$, $d = 0.179$). The distribution of calculated ViSQOL scores are visualized and grouped in Figure \ref{fig:boxplot-gender} which shows the differences in distributions per gender for each of the codecs and bitrates.

\vspace{2mm}
\noindent
\textbf{Age, dialect region and race.}
The DARPA-TIMIT corpus \cite{garofolo1993darpa} includes additional speaker metadata such as age, dialect region (8 major dialects of the USA), race and education level. As such, we also performed an N-way ANOVA with respect to each of these variables, followed by a post hoc analysis for each of the variables independently. The results indicate no biases of any audio codecs with respect to speaker age, dialect region, race or education level. For some codecs and bitrates, the results were not statistically significant (i.e. $p > 0.05$). For others, the results were statistically significant, but the observed effect sizes were either non-existent or too small; e.g., the largest observed partial eta squared was $\eta^{2} = 0.0184$ for G.729A, while most $\eta^{2}$ values were less than $0.01$.

%% file: tables/anova_t_test_language_group.tex
\begin{table}[!ht]
\tiny
\caption{Results of one-way ANOVA and t-test for each codec at a given bitrate between Germanic and Romance language groups. p-values have been corrected using the Benjamini-Hochberg procedure to adjust for multiple comparisons.}
\resizebox{\linewidth}{!}{%
\begin{tabular}{l|rrr|rrr}
\multicolumn{1}{c|}{}                     & \multicolumn{3}{c|}{ANOVA}                                                    & \multicolumn{3}{c}{t-test}                                                 \\ \hline
\multicolumn{1}{c|}{Codec (bitrate kbps)} & \multicolumn{1}{c}{F} & \multicolumn{1}{c}{p (corr)} & \multicolumn{1}{c|}{$\eta^{2}$} & \multicolumn{1}{c}{t} & \multicolumn{1}{c}{p (corr)} & \multicolumn{1}{c}{d} \\ \hline
EnCodec (6)                               & 4722                  & $ \ll $.0001                  & 0.0286                   & 68.71                 & $ \ll $.0001                  & 0.353                 \\
EnCodec (12)                              & 4525                  & $ \ll $.0001                  & 0.0275                   & 67.27                 & $ \ll $.0001                  & 0.345                 \\
EnCodec (24)                              & 4067                  & $ \ll $.0001                  & 0.0248                   & 63.78                 & $ \ll $.0001                  & 0.328                 \\
EnCodec (3)                               & 3980                  & $ \ll $.0001                  & 0.0243                   & 63.09                 & $ \ll $.0001                  & 0.324                 \\ \hline
Opus (24)                                 & 1748                  & $ \ll $.0001                  & 0.0090                   & 41.80                 & $ \ll $.0001                  & 0.196                 \\
Opus (12)                                 & 1395                  & $ \ll $.0001                  & 0.0086                   & 37.35                 & $ \ll $.0001                  & 0.192                 \\
Opus (8)                                  & 1327                  & $ \ll $.0001                  & 0.0082                   & 36.43                 & $ \ll $.0001                  & 0.187                 \\
Opus (6)                                  & 675                   & $ \ll $.0001                  & 0.0042                   & 25.99                 & $ \ll $.0001                  & 0.134                 \\ \hline
G.711A (64)                                 & 641                  & $ \ll $.0001                  & 0.0040                   & 25.20                 & $ \ll $.0001                  & 0.130                 \\
G.711$\mu$ (64)                                 & 623                   & $ \ll $.0001                  & 0.0039                   & 24.96                 & $ \ll $.0001                  & 0.128                 \\ \hline
G.729A (8)                                 & 248                   & $ \ll $.0001                  & 0.0015                   & 15.75                 & $ \ll $.0001                  & 0.081                 \\ \hline
Descript (8)                              & 58                    & $ \ll $.0001                  & 0.0004                   & 7.59                  & $ \ll $.0001                  & 0.039                
\end{tabular}
}
\label{tab::anova-t-test-language-groups}
\end{table}

%% file: text/discussion.tex
\section{Discussion}

Regarding language analysis, we obtained statistically significant results indicating that the difference of ViSQOL between the original and transcoded audio depend on the language of the audio. The effect size of this dependence varies between codecs and this can be inferred from Table \ref{tab::visqol-languages} as well. For example, for EnCodec at \SI{12}{kbps} and \SI{24}{kbps} the difference between Dutch and Spanish is about 10.2\%.
The language bias/sensitivity has a stronger effect for neural codecs (Descript excluded) compared to VoIP and PSTN. The results from Table \ref{tab::anova-t-test-language-groups} support this, as a medium effect ($\eta^{2} > 0.01$, $d > 0.3$) was observed only for EnCodec.

Our results suggest that neural codecs, particularly EnCodec, have a bias towards Romance languages compared to PSTN and VoIP codecs. Given that the highest disparity is between Dutch and Spanish, we hypothesize that neural codecs are more sensitive to phonological differences between languages compared to PSTN and VoIP codecs, i.e. neural codecs may have difficulties generalizing certain language constructs that they had not seen during training. As an example, in \cite{burgos13_interspeech} the authors report on a study on pronunciation errors by Spanish learners of Dutch, in which an overview of the phonological difference between Spanish and Dutch is also provided. Similar studies were also reported, all implying the relative difficulty of certain languages (in this case Dutch) and its phonological differences from other languages \cite{van2009automatic,doremalen2013automatic}.

\input{tables/gender_anova_t_test}

Regarding gender, we obtained statistically significant results which indicate that audio codecs are biased towards male voices compared to female. Contrary to the language differences in which neural codecs were most biased and PSTN codecs least affected, here we observed the reverse. PSTN codecs are significantly more biased towards male speech, with the means of the ViSQOL scores being more than a standard deviation apart ($d > 1$). Finally, our results are aligned with those of \cite{muller24c_interspeech} in which the authors obtained similar results for EnCodec; it has a bias towards male speech compared to female, especially at higher bitrates.

We hypothesize that the disparity in language and gender biases is expected; i.e. we argue that the higher language bias of neural codecs is explained by insufficiently diverse and unrepresentative data (in terms of language), while the higher gender bias of PSTN codecs is explained by the underlying algorithm of PSTN codecs being tailored towards spectra of male speech.
For example, Hillenbrand and Clark \cite{hillenbrand2009role} analyzed the role of the fundamental frequency F0 and referenced several studies which have shown a difference in F0 between men and women (slighthly less than an octave). More research reports similar results \cite{titze1989physiologic,kwon2010gender}. Given that PSTN codecs operate on predefined rules, we hypothesize that these rules do not equally model male and female speech, thereby resulting in large disparities between the two.

On the other hand, neural codecs are more dependent on the data which they are trained on. Given our results, this data is sufficiently balanced in terms of gender, but not sufficiently balanced in terms of language.

%% file: tables/gender_anova_t_test.tex
\begin{table}[t]
\tiny
\caption{Results of one-way ANOVA and t-test. A strong effect size of gender influencing ViSQOL is observable for PSTN codecs, while neural codecs at lower bitrates exhibit a neglible to small effect size.}
\resizebox{\linewidth}{!}{%
\begin{tabular}{l|rrr|rrr}
                     & \multicolumn{3}{c|}{ANOVA}                                                      & \multicolumn{3}{c}{t-test}                                                   \\ \hline
Codec (bitrate kbps) & \multicolumn{1}{c}{F} & \multicolumn{1}{c}{p (corr)} & \multicolumn{1}{c|}{$\eta^{2}$} & \multicolumn{1}{c}{t} & \multicolumn{1}{c}{p (corr)} & \multicolumn{1}{c}{d} \\ \hline
G.711mu (64)         & 2048                  & p $\ll$ .0001                      & 0.245                    & 54.55                 & p $\ll$ .0001                      & 1.239                 \\
G.711A (4)           & 1862                  & p $\ll$ .0001                      & 0.228                    & 52.41                 & p $\ll$ .0001                      & 1.181                 \\
G.729A (8)           & 1491                  & p $\ll$ .0001                      & 0.191                    & 51.81                 & p $\ll$ .0001                      & 1.057                 \\ \hline
Opus (8)             & 1016                  & p $\ll$ .0001                      & 0.139                    & 36.68                 & p $\ll$ .0001                      & 0.873                 \\
Opus (6)             & 864                   & p $\ll$ .0001                      & 0.121                    & 35.72                 & p $\ll$ .0001                      & 0.804                 \\
Opus (12)            & 495                   & p $\ll$ .0001                      & 0.073                    & 23.04                 & p $\ll$ .0001                      & 0.609                 \\ \hline
EnCodec (24)         & 184                   & p $\ll$ .0001                      & 0.028                    & 14.25                 & p $\ll$ .0001                      & 0.371                 \\
EnCodec (12)         & 139                   & p $\ll$ .0001                      & 0.022                    & 12.36                 & p $\ll$ .0001                      & 0.323                 \\
Descript (8)         & 117                   & p $\ll$ .0001                      & 0.018                    & 11.25                 & p $\ll$ .0001                      & 0.296                 \\ \hline
Opus (24)            & 71                    & p $\ll$ .0001                      & 0.011                    & 8.88                  & p $\ll$ .0001                      & 0.231                 \\
EnCodec (6)          & 43                    & p $\ll$ .0001                      & 0.007                    & 6.80                  & p $\ll$ .0001                      & 0.179                 \\
EnCodec (3)          & 9                     & p = .0032                    & 0.001                    & 3.04                  & p = .0024                    & 0.081                
\end{tabular}
}
\label{tab::anova-t-test-gender}
\end{table}

%% file: text/conclusion.tex
\section{Conclusion}

In this paper, we investigate traditional, VoIP-based and neural audio codecs with respect to language and geneder bias.
Our results showed that neural codecs have a higher language bias compared to the frequently used PSTN and VoIP codecs, with medium effect sizes observable between Germanic and Romance language groups. 
With respect to gender bias, our experiments showed a strong effect size; PSTN codecs create a significant disparity of audio quality between male and female speech. We provided plausible explanations for these phenomena supported by studies which report difficulties observed in second language learners given the phonological differences of their native language and the language that they are studying.

Overall, our results indicate that neural codecs should better adjust for the phonological differences between languages during training. For future work, we want to investigate how PSTN codecs can be corrected to compensate for the biases they introduce and to explore whether these biases persist in cases of poor network quality (e.g. packet loss or jitter).

%% file: template.bbl
\begin{thebibliography}{10}
\providecommand{\url}[1]{#1}
\csname url@samestyle\endcsname
\providecommand{\newblock}{\relax}
\providecommand{\bibinfo}[2]{#2}
\providecommand{\BIBentrySTDinterwordspacing}{\spaceskip=0pt\relax}
\providecommand{\BIBentryALTinterwordstretchfactor}{4}
\providecommand{\BIBentryALTinterwordspacing}{\spaceskip=\fontdimen2\font plus
\BIBentryALTinterwordstretchfactor\fontdimen3\font minus \fontdimen4\font\relax}
\providecommand{\BIBforeignlanguage}[2]{{%
\expandafter\ifx\csname l@#1\endcsname\relax
\typeout{** WARNING: IEEEtran.bst: No hyphenation pattern has been}%
\typeout{** loaded for the language `#1'. Using the pattern for}%
\typeout{** the default language instead.}%
\else
\language=\csname l@#1\endcsname
\fi
#2}}
\providecommand{\BIBdecl}{\relax}
\BIBdecl

\bibitem{gersho2002advances}
A.~Gersho, ``Advances in speech and audio compression,'' \emph{Readings in multimedia computing and networking}, pp. 23--41, 2002.

\bibitem{defossezhigh}
A.~D{\'e}fossez, J.~Copet, G.~Synnaeve, and Y.~Adi, ``High fidelity neural audio compression,'' \emph{Transactions on Machine Learning Research}, 2023.

\bibitem{g711}
``{ITU-T} {R}ecommendation {G}.711: {P}ulse code modulation ({PCM}) of voice frequencies,'' {N}ov 1988.

\bibitem{g729}
``{ITU-T} {R}ecommendation {G}.729: Coding of speech at 8kbit/s using conjugate-structure algebraic-code-excited linear prediction {CS-ACELP},'' {M}ar 1996.

\bibitem{dietz2015overview}
M.~Dietz, M.~Multrus, V.~Eksler, V.~Malenovsky, E.~Norvell, H.~Pobloth, L.~Miao, Z.~Wang, L.~Laaksonen, A.~Vasilache \emph{et~al.}, ``Overview of the evs codec architecture,'' in \emph{2015 IEEE International Conference on Acoustics, Speech and Signal Processing (ICASSP)}.\hskip 1em plus 0.5em minus 0.4em\relax IEEE, 2015, pp. 5698--5702.

\bibitem{valin2012definition}
J.-M. Valin, K.~Vos, and T.~Terriberry, ``Rfc 6716: Definition of the opus audio codec,'' 2012.

\bibitem{zeghidour2021soundstream}
N.~Zeghidour, A.~Luebs, A.~Omran, J.~Skoglund, and M.~Tagliasacchi, ``Soundstream: An end-to-end neural audio codec,'' \emph{IEEE/ACM Transactions on Audio, Speech, and Language Processing}, vol.~30, pp. 495--507, 2021.

\bibitem{kumar2024high}
R.~Kumar, P.~Seetharaman, A.~Luebs, I.~Kumar, and K.~Kumar, ``High-fidelity audio compression with improved rvqgan,'' \emph{Advances in Neural Information Processing Systems}, vol.~36, 2024.

\bibitem{tatman2017gender}
R.~Tatman, ``Gender and dialect bias in youtube’s automatic captions,'' in \emph{Proceedings of the first ACL workshop on ethics in natural language processing}, 2017, pp. 53--59.

\bibitem{tatman17_interspeech}
R.~Tatman and C.~Kasten, ``Effects of talker dialect, gender \& race on accuracy of bing speech and youtube automatic captions,'' in \emph{Interspeech 2017}, 2017, pp. 934--938.

\bibitem{koenecke2020racial}
A.~Koenecke, A.~Nam, E.~Lake, J.~Nudell, M.~Quartey, Z.~Mengesha, C.~Toups, J.~R. Rickford, D.~Jurafsky, and S.~Goel, ``Racial disparities in automated speech recognition,'' \emph{Proceedings of the national academy of sciences}, vol. 117, no.~14, pp. 7684--7689, 2020.

\bibitem{mengesha2021don}
Z.~Mengesha, C.~Heldreth, M.~Lahav, J.~Sublewski, and E.~Tuennerman, ``“i don’t think these devices are very culturally sensitive.”—impact of automated speech recognition errors on african americans,'' \emph{Frontiers in Artificial Intelligence}, vol.~4, p. 725911, 2021.

\bibitem{liu2022towards}
C.~Liu, M.~Picheny, L.~Sar{\i}, P.~Chitkara, A.~Xiao, X.~Zhang, M.~Chou, A.~Alvarado, C.~Hazirbas, and Y.~Saraf, ``Towards measuring fairness in speech recognition: Casual conversations dataset transcriptions,'' in \emph{ICASSP 2022-2022 IEEE International Conference on Acoustics, Speech and Signal Processing (ICASSP)}.\hskip 1em plus 0.5em minus 0.4em\relax IEEE, 2022, pp. 6162--6166.

\bibitem{dheram22_interspeech}
P.~DHERAM, M.~Ramakrishnan, A.~Raju, I.-F. Chen, B.~King, K.~Powell, M.~Saboowala, K.~Shetty, and A.~Stolcke, ``Toward fairness in speech recognition: Discovery and mitigation of performance disparities,'' in \emph{Interspeech 2022}, 2022, pp. 1268--1272.

\bibitem{gorrostieta19_interspeech}
C.~Gorrostieta, R.~Lotfian, K.~Taylor, R.~Brutti, and J.~Kane, ``Gender de-biasing in speech emotion recognition,'' in \emph{Interspeech 2019}, 2019, pp. 2823--2827.

\bibitem{luitel2024investigating}
S.~Luitel, Y.~Liu, and M.~Anwar, ``Investigating fairness in machine learning-based audio sentiment analysis,'' \emph{AI and Ethics}, pp. 1--10, 2024.

\bibitem{holub2023human}
J.~Holub and Y.~Kowalczuk, ``Human-centered design of voice communications: Gender aspects,'' \emph{Human Interaction and Emerging Technologies (IHIET-AI 2023): Artificial Intelligence and Future Applications}, vol.~70, no.~70, 2023.

\bibitem{lai24_interspeech}
L.-F. Lai and N.~Holliday, ``Voice quality variation in aae: An additional challenge for addressing bias in asr models?'' in \emph{Interspeech 2024}, 2024, pp. 3080--3084.

\bibitem{xu24j_interspeech}
C.~Xu, J.~Wormald, P.~Foulkes, P.~Harrison, V.~Hughes, P.~Welch, F.~Kelly, and D.~{van der Vloed}, ``Voice quality in telephone speech: Comparing acoustic measures between voip telephone and high-quality recordings,'' in \emph{Interspeech 2024}, 2024, pp. 1570--1574.

\bibitem{muller24c_interspeech}
T.~Muller, S.~Ragot, L.~Gros, P.~Philippe, and P.~Scalart, ``Speech quality evaluation of neural audio codecs,'' in \emph{Interspeech 2024}, 2024, pp. 1760--1764.

\bibitem{ardila2020common}
R.~Ardila, M.~Branson, K.~Davis, M.~Kohler, J.~Meyer, M.~Henretty, R.~Morais, L.~Saunders, F.~Tyers, and G.~Weber, ``Common voice: A massively-multilingual speech corpus,'' in \emph{Proceedings of the Twelfth Language Resources and Evaluation Conference}, 2020, pp. 4218--4222.

\bibitem{dubey2024icassp}
H.~Dubey, A.~Aazami, V.~Gopal, B.~Naderi, S.~Braun, R.~Cutler, A.~Ju, M.~Zohourian, M.~Tang, M.~Golestaneh \emph{et~al.}, ``Icassp 2023 deep noise suppression challenge,'' \emph{IEEE Open Journal of Signal Processing}, 2024.

\bibitem{mysore2014can}
G.~J. Mysore, ``Can we automatically transform speech recorded on common consumer devices in real-world environments into professional production quality speech?—a dataset, insights, and challenges,'' \emph{IEEE Signal Processing Letters}, vol.~22, no.~8, pp. 1006--1010, 2014.

\bibitem{VoxForge}
\BIBentryALTinterwordspacing
{VoxForge}, ``{VoxForge: Free and Open Source Speech Dataset},'' 2006, accessed: 2024-12-16. [Online]. Available: \url{http://www.voxforge.org/}
\BIBentrySTDinterwordspacing

\bibitem{garofolo1993darpa}
J.~S. Garofolo, L.~F. Lamel, W.~M. Fisher, J.~G. Fiscus, and D.~S. Pallett, ``Darpa timit acoustic-phonetic continous speech corpus cd-rom. nist speech disc 1-1.1,'' \emph{NASA STI/Recon technical report n}, vol.~93, p. 27403, 1993.

\bibitem{chinen2020visqol}
M.~Chinen, F.~S. Lim, J.~Skoglund, N.~Gureev, F.~O'Gorman, and A.~Hines, ``Visqol v3: An open source production ready objective speech and audio metric,'' in \emph{2020 twelfth international conference on quality of multimedia experience (QoMEX)}.\hskip 1em plus 0.5em minus 0.4em\relax IEEE, 2020, pp. 1--6.

\bibitem{rfc6716}
\BIBentryALTinterwordspacing
J.-M. Valin, K.~Vos, and T.~B. Terriberry, ``{Definition of the Opus Audio Codec},'' RFC 6716, Sep. 2012. [Online]. Available: \url{https://www.rfc-editor.org/info/rfc6716}
\BIBentrySTDinterwordspacing

\bibitem{tibshirani1993introduction}
R.~J. Tibshirani and B.~Efron, ``An introduction to the bootstrap,'' \emph{Monographs on statistics and applied probability}, vol.~57, no.~1, pp. 1--436, 1993.

\bibitem{benjamini1995controlling}
Y.~Benjamini and Y.~Hochberg, ``Controlling the false discovery rate: a practical and powerful approach to multiple testing,'' \emph{Journal of the Royal statistical society: series B (Methodological)}, vol.~57, no.~1, pp. 289--300, 1995.

\bibitem{ellis2010essential}
P.~D. Ellis, \emph{The essential guide to effect sizes: Statistical power, meta-analysis, and the interpretation of research results}.\hskip 1em plus 0.5em minus 0.4em\relax Cambridge university press, 2010.

\bibitem{burgos13_interspeech}
P.~Burgos, C.~Cucchiarini, R.~van Hout, and H.~Strik, ``Pronunciation errors by spanish learners of dutch: a data-driven study for asr-based pronunciation training,'' in \emph{Interspeech 2013}, 2013, pp. 2385--2389.

\bibitem{van2009automatic}
J.~Van~Doremalen, C.~Cucchiarini, and H.~Strik, ``Automatic detection of vowel pronunciation errors using multiple information sources,'' in \emph{2009 IEEE Workshop on Automatic Speech Recognition \& Understanding}.\hskip 1em plus 0.5em minus 0.4em\relax IEEE, 2009, pp. 580--585.

\bibitem{doremalen2013automatic}
J.~v. Doremalen, C.~Cucchiarini, and H.~Strik, ``Automatic pronunciation error detection in non-native speech: The case of vowel errors in dutch,'' \emph{The Journal of the Acoustical Society of America}, vol. 134, no.~2, pp. 1336--1347, 2013.

\bibitem{hillenbrand2009role}
J.~M. Hillenbrand and M.~J. Clark, ``The role of f 0 and formant frequencies in distinguishing the voices of men and women,'' \emph{Attention, Perception, \& Psychophysics}, vol.~71, pp. 1150--1166, 2009.

\bibitem{titze1989physiologic}
I.~R. Titze, ``Physiologic and acoustic differences between male and female voices,'' \emph{The Journal of the Acoustical Society of America}, vol.~85, no.~4, pp. 1699--1707, 1989.

\bibitem{kwon2010gender}
H.-B. Kwon, ``Gender difference in speech intelligibility using speech intelligibility tests and acoustic analyses,'' \emph{The journal of advanced prosthodontics}, vol.~2, no.~3, p.~71, 2010.

\end{thebibliography}
